\documentclass[aps,preprint,nofootinbib]{revtex4}
\usepackage{amsfonts}
\usepackage{amsmath}
\usepackage{amssymb}
\usepackage{graphicx}

\providecommand{\U}[1]{\protect\rule{.1in}{.1in}}

\input{tcilatex}

\begin{document}

\preprint{}
\title[ ]{Hyperfine structure and $(e^{-},e^{+})$-pair annihilation in the
muonium-positronium MuPs and positronium hydrides.}
\author{Alexei M. Frolov }
\email{afrolov@uwo.ca}
\author{David M. Wardlaw}
\affiliation{Department of Chemistry, University of Western Ontario, London, Canada}
\date{25 November 2009}

\begin{abstract}
The hyperfine structure of the ground states in a number of positronium
hydrides (TPs, DPs, ${}^{1}$HPs) and MuPs ($\mu ^{+}e_{2}^{-}e^{+}$) is
determined with the use of highly accurate variational wave functions. We
also evaluate the probabilities of various processes in the MuPs system,
including the $(e^{-},e^{+})$-pair annihilation and its conversion into the
charge conjugate system $\mu ^{+}e_{2}^{-}e^{+}\rightarrow \mu
^{-}e^{-}e_{2}^{+}$.
\end{abstract}

\maketitle

\vspace{0.5cm}

\section{Introduction.}

In this work we consider the bound states in the positronium hydrides $%
{}^{\infty}$HPs, TPs, DPs, ${}^{1}$HPs and MuPs ($\mu^{+} e^{-}_2 e^{+}$).
The last system is the main interest in this study. Each of these neutral
systems contain one heavy positively charged particle, i.e. hydrogen nucleus
or $\mu^{+}$, two electrons $e^{-}$ and one positron $e^{+}$. Below, such
four-body systems are designated as $A^{+} e^{-}_2 e^{+}$ systems, where the
notation $A$ designates a heavy particle with $m_A \gg m_e$. In atomic units 
$\hbar = 1, m_e = 1, e = 1$ the Hamiltonian of the four-body $A^{+} e^{-}_2
e^{+}$ system is written in the form (in atomic units): 
\begin{eqnarray}
H = -\frac{1}{2 m_{A}} \Delta_{1} -\frac{1}{2} \Delta_{2} -\frac{1}{2}
\Delta_{3} -\frac{1}{2} \Delta_{4} + \frac{1}{r_{12}} - \frac{1}{r_{13}} - 
\frac{1}{r_{14}} - \frac{1}{r_{23}} - \frac{1}{r_{24}} + \frac{1}{r_{34}}
\label{eq1}
\end{eqnarray}
where the notation 1 (also $A$) designates the heaviest particle $A^{+}$,
the notation 2 (or +) means the positron, while 3 (or -) and 4 (or -) stand
for electrons. This system of notations will be used everywhere below in our
study.

Our first goal is to determine the wave functions which correspond to the
bound states of the Hamiltonian, Eq.(\ref{eq1}). In other words, we need to
find all negative eigenvalues $E$ and unit-norm functions $\Psi$ for which
the corresponding Schr\"{o}dinger equation $H \Psi = E \Psi$ is obeyed. It
is clear that the total energies and other bound state properties of the $%
A^{+} e^{-}_2 e^{+}$ systems must be analytical functions of the inverse
masses of heavy particle $\frac{1}{m_A}$. Note that some state in the $A^{+}
e^{-}_2 e^{+}$ system is stable if its total energy (in atomic units) is
less than the corresponding threshold value 
\begin{eqnarray}
E_{th} (m_{A}) = -\frac{0.5}{1 + m_{A}^{-1}} - 0.25 = -\frac{(3 m_A + 1)}{4
\cdot (m_A + 1)} \geq -\frac{3}{4} = -0.75 a.u.
\end{eqnarray}
As follows from the results of numerical calculations for all positronium
hydrides the total energy of the ground state is bounded between $\approx$
-0.78631730(15) $a.u.$ (MuPs) and $\approx$ -0.789196770(3) $a.u.$ ($%
{}^{\infty}$HPs). This indicates clearly that each of these positronium
hydrides is a weakly bound four-body system. In fact, it was shown long ago
that each of these hydrides has only one bound (ground) $S-$state \cite{Ore}%
, where $S$ designates a state with $L = 0$ and $L$ is the total orbital
angular momentum of the four-body system. Moreover, it is bound if (and only
if) the two electrons form the singlet pair, i.e. the total electron spin
equals zero.

In general, the positronium hydrides are of interest for astrophysics \cite%
{Dra1}, \cite{Dra2}. Almost 20 years ago the ${}^{1}$HPs hydride was created
in the laboratory during collisions between positrons and methane \cite{HPs}%
. Theoretically, the positronium hydride ${}^{\infty}$HPs has extensively
been investigated in earlier studies \cite{Hou} - \cite{Sch}, \cite{Fro1}, 
\cite{Bubin} (all references on HPs before 1998 can be found in \cite{FrSm}%
). The bound muonium-positronium MuPs has never been observed in the
laboratory.

In this work our main attention will be given to some properties of the
muonium-positronium system (or MuPs, for short), but we also evaluate the
probabilities of some processes within it. The hyperfine structure of the
MuPs system is discussed in Section II. Section III contains numerical
evaluations of different annihilation probabilities for muonium-positronium.
In Section IV we consider the annihilation rates of the electron-positron
pairs in other positronium hydrides. In Section V we discuss a possibility
to observe the conversion of MuPs into its charge conjugate system $\mu^{+}
e^{-}_2 e^{+} \rightarrow \mu^{-} e^{-} e^{+}_2$. Concluding remarks can be
found in Section VI.

\section{The hyperfine structure of the ground state in muonium-positronium.}

The hyperfine structure (i.e. the appropriate shift of the energy level and
its splitting) is determined by the spin-spin interaction between particles.
The general expression for the hyperfine interaction of a number of
particles with non-zero spin values can be written in the form 
\begin{equation}
H_{HF} = - \sum_{(ij)} a_{ij} (\mathbf{s}_{i} \cdot \mathbf{s}_{j})
\label{spl1}
\end{equation}
where in the case of $A^{+} e^{-}_2 e^{+}$ system the sum is calculated for
all six pairs of particles $(ij)$. However, as mentioned above in the ground
state of the MuPs system the two electrons are always in the singlet state,
i.e. their total spin equals zero. Also, in this work we are interested in
the hyperfine structure splitting only. In such a case Eq.(\ref{spl1}) can
be re-written to the form 
\begin{eqnarray}
H_{HF} = -a (\mathbf{I}_{A} \cdot \mathbf{s}_{+}) - b (\mathbf{s}_{+} \cdot 
\mathbf{S}_{-}) - c (\mathbf{I}_{A} \cdot \mathbf{S}_{-})  \label{spl2}
\end{eqnarray}
where $\mathbf{S}_{-}$ is the total electron spin (i.e. $\mathbf{S}_{-} = 
\mathbf{s}_1 + \mathbf{s}_2$ in our current notations), $\mathbf{s}_{+}$ is
the positron spin and $\mathbf{I}_{A}$ is the spin of the $A-$particle.

In the MuPs system both electrons are in the singlet state, i.e. $\mathbf{S}%
_{-} = 0$. Therefore, from Eq.(\ref{spl2}) one finds $H_{HF} = -a (\mathbf{I}%
_{\mu} \cdot \mathbf{s}_{+})$, where the coupling constant $a$ is written in
the form 
\begin{equation}
a = \frac{8 \pi \alpha^{2}}{3} \mu_{B}^{2} \cdot \frac{g_{\mu}}{m_{\mu}}
\cdot \frac{g_{+}}{m_e} \cdot \langle \delta_{\mu^{+} e^{+}} \rangle = \frac{%
2 \pi \alpha^{2}}{3} \cdot \frac{g_{\mu}}{m_{\mu}} \cdot g_{+} \cdot \langle
\delta_{\mu^{+} e^{+}} \rangle  \label{spl3}
\end{equation}
where $\langle \delta_{\mu^{+} e^{+}} \rangle$ is the expectation value of
the muon-positron delta-function $\delta_{\mu^{+} e^{+}} = \delta(\mathbf{r}%
_{\mu^{+} e^{+}}) = \delta(\mathbf{r}_{\mu^{+}} - \mathbf{r}_{e^{+}})$
determined for the ground state of MuPs and expressed in atomic units. Also,
In Eq.(\ref{spl3}) the factor $\alpha = 7.2973525679 \cdot 10^{-3}$ is the
fine structure constant and $\mu_{B} = \frac{e \cdot \hbar} {2 m_{e}}$ is
the Bohr magneton which equals $\frac{1}{2}$ in the atomic units ($e = 1,
\hbar = 1$ and $m_{e} = 1$). The value of $\mu_{B}$ in SI units is $\approx
9.27401543 \cdot 10^{-24} J \cdot T^{-1}$ \cite{NIST}. In our calculations
we have used the following values for the muon mass $m_{\mu}$ and for the
factors $g_{+}$ and $g_{\mu}$ \cite{NIST}, \cite{CRC}: 
\begin{eqnarray}
m_{\mu} = 206.768264 m_e \; \; \; , \; \; \; g_{+} = -2.0023193043718 \; \;
\; \; \; \; g_{\mu} = -2.0023318396
\end{eqnarray}
where $m_e$ is the electron/positron mass at rest. With these numerical
values Eq.(\ref{spl3}) takes the form 
\begin{equation}
a = 14229.1255 \cdot \langle \delta_{\mu^{+} e^{+}} \rangle  \label{spl31}
\end{equation}

For MuPs the diagonalization of the $H_{HF}$ operator yields the two
energies: $\epsilon(J=0) = \frac34 a$ and $\epsilon(J=1) = -\frac14 a$,
where $a > 0$. The notation $J$ denotes the total spin of the muon-positron
pair. From our numerical calculations we have found that in the ground state
of MuPs the numerical value of muon-positron delta-function is $\langle
\delta_{\mu^{+} +} \rangle \approx 1.613451 \cdot 10^{-3}$. From here one
finds that the energy difference between $\epsilon(J=0) = \frac34 a$ and $%
\epsilon(J=1) = -\frac14 a$ levels equals $a \approx$ 22.985 $MHz$. The
uncertainty in this value can be evaluated as $\approx$ 10 $kHz$. To convert
the atomic units into $MHz$ we have used the conversion factor 6.57968392061$%
\cdot 10^9$ $MHz / a.u$. The value 22.958 $MHz$ must be compared with the
total ground state energy (non-relativistic) obtained for the MuPs system $E$
= -0.7683171715 $a.u.$ $\approx$ 5.0552841393$\cdot 10^{9}$ $MHz$. Analogous
calculations of the hyperfine structure splitting can be performed for all
positronium hydrides mentioned above (see Section IV below).

\section{Electron-positron annihilation in muonium-positronium.}

The muonium-positronium system is not a stable four-body system. Its
instability is mainly related with the $(e^{-},e^{+})-$pair annihilation. In
some works such an annihilation is called the positron annihilation. The
life-time of MuPs against positron annihilation is $\approx 2.247 \cdot
10^{-10}$ $sec$ (see below). Another possible decay channel arises from the
instability of the $\mu^{+}$-muon. It usually decays into one positron, one
electron neutrino and one muon antineutrino (see Section IV below). The
corresponding life-time is $\approx$ 2.19703 $(\pm 4 \cdot 10^{-5}) \cdot
10^{-6}$ $sec$ which is approximately 15 times longer than the life-time of
MuPs against three-photon annihilation. Muonium-positronium conversion in
MuPs is also possible (see discussion in the fourth Section). In this
Section we consider annihilation of the $(e^{-},e^{+})-$pair in the ground
state of the MuPs system.

First, consider the two- and three-photon annihilation rates. As is well
known from Quantum Electrodynamics (see, e.g., \cite{AB}) an isolated
electron-positron pair or Ps ($e^{-} e^{+}$), which is in the singlet ${}^1S-
$state, annihilates with the emission of two, four, six and any even number
of photons. The largest annihilation rate is for two-photon annihilation: 
\begin{eqnarray}
\Gamma_{2 \gamma}(\mathrm{Ps}, {}^1S) = 4 \pi \alpha^4 c a^{-1}_0 \Bigl[ 1 - 
\frac{\alpha}{\pi} \Bigl( 5 - \frac{\pi^2}{4} \Bigr)\Bigr] \langle \delta(%
\mathbf{r}_{+-}) \rangle = 4 \times 50.17280269804 \cdot 10^{9} \cdot
\langle \delta_{+-} \rangle \; sec^{-1} \; \; \; ,  \label{ep2}
\end{eqnarray}
where the notation $\delta(\mathbf{r}_{+-}) = \delta_{+-}$ is the two-body
electron-positron delta-function and $\langle \delta_{+-} \rangle$ is its
expectation value determined for the singlet ${}^1S-$state of
electron-positron pair. In this formula and everywhere below we shall use
the following numerical values for speed of light $c = 2.99792458 \cdot
10^{8}$ $m \cdot sec^{-1}$ and for Bohr radius $a_0 = 0.5291772108 \cdot
10^{-10}$ $m$ \cite{NIST}. Note that our expression for $\Gamma_{2 \gamma}$,
Eq.(\ref{ep2}), also includes the lowest order radiative correction \cite%
{Harr}. Analogously, an isolated electron-positron pair, which is in the
triplet ${}^3S-$state, annihilates with the emission of three, five, seven
and any odd number of photons. The largest annihilation rate is for
three-photon annihilation: 
\begin{eqnarray}
\Gamma_{3 \gamma}(\mathrm{Ps}, {}^3S) = \frac{16 (\pi^2 - 9)}{9} \alpha^5 c
a^{-1}_0 \langle \delta(\mathbf{r}_{+-}) \rangle = \frac43 \times
1.35927229774 \cdot 10^8 \langle \delta_{+-} \rangle \; sec^{-1} \; \; \; .
\label{ep3}
\end{eqnarray}

In an arbitrary atom, ion or molecule which contain the bound positron we
have a number of electron-positron pairs which are generally in mixed spin
states and this is the case in the MuPs system. This means that we cannot
predict the actual spin state of these electron-positron pairs. In such
cases it is assumed that each of the four possible spin states of the
electron-positron pair has equal probability, which implies the probability
of $\frac14$ to be in its singlet state and the probability of $\frac34$ to
be in its triplet state \cite{Heitl}. The total probability of the
two-photon annihilation of $(e^{-},e^{+})-$pair which is in a mixed spin
state equals the product of $\Gamma_{2 \gamma}(\mathrm{Ps}, {}^1S)$ (Eq.(\ref%
{ep2})), the factor $\frac14$, and the number of electron-positron pairs $n$%
. In this case one finds the formulae presented above for the MuPs system ($n
$ = 2) 
\begin{eqnarray}
\Gamma_{2 \gamma}(\mathrm{MuPs}) = n \pi \alpha^4 c a^{-1}_0 \Bigl[ 1 - 
\frac{\alpha}{\pi} \Bigl( 5 - \frac{\pi^2}{4} \Bigr)\Bigr] \langle \delta(%
\mathbf{r}_{+-}) \rangle = 100.3456053781 \cdot 10^{9} \langle \delta_{+-}
\rangle \; sec^{-1} \; \; \; .  \label{An2g}
\end{eqnarray}
In the case of three-photon annihilation the $\Gamma_{3 \gamma}(\mathrm{MuPs}%
)$ annihilation rate equals the product of $\Gamma_{3 \gamma}(\mathrm{Ps},
{}^3S)$ (Eq.(\ref{ep3})), the factor $\frac34$, and the number of
electron-positron pairs $n$, i.e. 
\begin{eqnarray}
\Gamma_{3 \gamma}(\mathrm{MuPs}) = n \frac{4 (\pi^2 - 9)}{3} \alpha^5 c
a^{-1}_0 \langle \delta(\mathbf{r}_{+-}) \rangle = 2.718545954 \cdot 10^8
\langle \delta_{+-} \rangle \; sec^{-1} \; \; \; ,
\end{eqnarray}
where $\langle \delta_{+-} \rangle$ is the expectation value of the
electron-positron delta-function determined for the ground state in the MuPs
system.

Now, let us discuss the four- and five-photon annihilation of the
electron-positron pairs in the MuPs system. It was shown in \cite{PRA83}
that the rates of the four- and two-photon annihilation in para-positronium
(i.e. in the $(e^{-}, e^{+}$)-pair in its singlet state) are related to each
other by the following approximate equation 
\begin{equation}
\Gamma_{4 \gamma}(\mathrm{Ps}, {}^1S) \approx 0.274 \Bigl(\frac{\alpha}{\pi}%
\Bigr)^2 \Gamma_{2 \gamma}(\mathrm{Ps}, {}^1S)  \label{e4}
\end{equation}
By multiplying the both sides of this equation by the factor $\frac14$ and
the total number of electron-positron pairs (in MuPs $n = 2$) one finds an
analogous expression for the MuPs system 
\begin{equation}
\Gamma_{4 \gamma}(\mathrm{MuPs}) \approx 0.274 \Bigl(\frac{\alpha}{\pi}\Bigr)%
^2 \Gamma_{2 \gamma}(\mathrm{MuPs})  \label{e4a}
\end{equation}
where $\Gamma_{4 \gamma}$(MuPs) and $\Gamma_{2 \gamma}$(MuPs) are the
corresponding annihilation rates of the MuPs system. For the two-photon
annihilation rate $\Gamma_{2 \gamma}$ in Eq.(\ref{e4a}) one can use the
explicit expression Eq.(\ref{An2g}). Note that in Eq.(\ref{e4a}) the formula
for the $\Gamma_{2 \gamma}$ rate must be used which does not contain the
lowest order radiative correction. But, for approximate evaluations we can
ignore such a small difference in $\Gamma_{2 \gamma}$. For the five-photon
annihilation rate in the MuPs system one analogously finds the following
result 
\begin{equation}
\Gamma_{5 \gamma}(\mathrm{MuPs}) \approx 0.177 \Bigl(\frac{\alpha}{\pi}\Bigr)%
^2 \Gamma_{3 \gamma}(\mathrm{MuPs})  \label{e5a}
\end{equation}
This result is based on the formula from Ref.\cite{PRA83}. The numerical
values of the $\Gamma_{2 \gamma}, \Gamma_{3 \gamma}, \Gamma_{4 \gamma}$ and $%
\Gamma_{5 \gamma}$ annihilation rates computed with the use of these
formulas are: $\Gamma_{2 \gamma} \approx 2.4522354(30) \cdot 10^9$ $sec^{-1}$%
, $\Gamma_{3 \gamma} \approx 6.643554(10) \cdot 10^6$ $sec^{-1}$, $\Gamma_{4
\gamma} \approx 3.6253(1) \cdot 10^3$ $sec^{-1}$ and $\Gamma_{5 \gamma}
\approx 6.1723(1)$ $sec^{-1}$. They also can be found in Table I. The four-
and five-photon annihilation rates (i.e. $\Gamma_{4 \gamma}$ and $\Gamma_{5
\gamma}$) have never been evaluated (accurately) in earlier studies. Table I
also contains the numerical values of $\Gamma_{2 \gamma}, \Gamma_{3 \gamma},
\Gamma_{4 \gamma}, \Gamma_{5 \gamma}$ annihilation rates determined for the
positronium hydrides ${}^{\infty}$HPs, TPs, DPs and ${}^{1}$HPs. The
numerical values of these $n-$photon annihilation rates allow one to
estimate the total annihilation rate $\Gamma \approx \Gamma_{2 \gamma} +
\Gamma_{3 \gamma} + \Gamma_{4 \gamma} + \Gamma_{5 \gamma} \approx \Gamma_{2
\gamma} + \Gamma_{3 \gamma}$ for each of the positronium hydrides and MuPs.

The two-, three-, four- and five-photon annihilations are the leading
annihilation processes in MuPs and other positronium hydrides. In some
applications, however, the one-photon and zero-photon annihilations may also
play an important role. For the zero-photon annihilation rate $\Gamma_{0
\gamma}$ we shall use the following expression (found in \cite{Kru}) 
\begin{eqnarray}
\Gamma_{0 \gamma} = \xi \frac{147 \sqrt{3} \pi^3}{2} \cdot \alpha^{12} (c
a_0^{-1}) \cdot \langle \delta_{\mu^{+}+--} \rangle = 5.0991890 \cdot
10^{-4} \cdot \xi \cdot \langle \delta_{\mu^{+}+--} \rangle \; \; \; sec^{-1}
\end{eqnarray}
where $\langle \delta_{\mu^{+}+--} \rangle$ is the expectation value of the
four-particle delta-function in the ground state of muonium-positronium
(MuPs). Its numerical value is the probability to find all four particles at
one spatial point with spatial radius $\alpha a_0$. The unknown
(dimensionless) factor $\xi$ has the numerical value close to unity. The
expectation value of the four-particle delta-function determined in our
calculations is $\approx 1.785222 \cdot 10^{-4}$ (in $a.u.$). From here one
finds that $\Gamma_{0 \gamma}$(MuPs) $\approx 9.10318(10) \cdot 10^{-8} \xi$ 
$sec^{-1}$. For approximate evaluations we can assume that the factor $\xi$
equals unity. In this case one finds that $\Gamma_{0 \gamma}$(MuPs) $\approx
9.10318(10) \cdot 10^{-8}$ $sec^{-1}$.

Now, consider the one-photon annihilation of the electron-positron pair in
MuPs (this can proceed with the emission of one fast electron). The
probability of such a process is given by the formula (its rigorous
derivation can be found in \cite{Kry}) 
\begin{eqnarray}
\Gamma^{(1)}_{1 \gamma} = \frac{64 \pi^2}{27} \cdot \alpha^{8} (c a_0^{-1})
\cdot \langle \delta_{+--} \rangle = 1.066420947 \cdot 10^3 \cdot \langle
\delta_{+--} \rangle \; \; \; sec^{-1} ,
\end{eqnarray}
where $\langle \delta_{+--} \rangle = \langle \delta(\mathbf{r}_{+-}) \delta(%
\mathbf{r}_{--}) \rangle)$ is the expectation value of the triple
electron-positron delta-function in the ground state of the MuPs system. Its
numerical value is the probability to find all three corresponding particles
at one spatial point with spatial radius $\alpha a_0$. Our best numerical
treatment to-date gives $\langle \delta_{+--} \rangle \approx 3.631815 \cdot
10^{-4}$ resulting in $\Gamma^{(1)}_{1 \gamma} \approx$ 3.87063(10)$\cdot
10^{-1}$ $sec^{-1}$ for the MuPs ground state.

In addition to this one-photon annihilation in MuPs another one-photon
annihilation of the $(e^{-}, e^{+})$-pair is possible. In \cite{FrSm} such
an annihilation was called the second one-photon annihilation. The
corresponding annihilation rate is designated as $\Gamma^{(2)}_{1 \gamma}$.
In this case the probability of one-photon annihilation is $\sim \langle
\delta_{\mu^{+}+-} \rangle$ and one (of two) annihilation $\gamma-$quanta is
absorbed by the heavy $\mu^{+}$ muon. The muon takes all photon's energy
(i.e. $\approx$ 0.51099906 $MeV$) and its momentum. The Lorentz $\gamma-$%
factor of the final/accelerated muon is $\approx$ 1.00483633, i.e. the
acceleration of the final $\mu^{+}$ muon produced by the absorbed $\gamma-$%
quantum is very small. It follows from here that the final $\mu^{+} e^{-}$
system (or muonium) can be found either in its ground $1S(L = 0)-$state ($%
P_g \ge 97 \%$), or in the excited $2S(L = 0)-$state ($P_e \le 1 \%$), or in
the unbound state ($P_u \le 1 \%$). Formally, the rate of the second
one-photon annihilation in the MuPs system can be evaluated from the
approximate equality $\Gamma^{(2)}_{1 \gamma} \approx \Gamma^{(1)}_{1 \gamma}
$. To obtain the more accurate value of $\Gamma^{(2)}_{1 \gamma}$ one needs
undertake an extensive QED consideration.

\section{Hyperfine structure and electron-positron annihilation in
positronium hydrides.}

The formulas presented above can also be used to compute the hyperfine
structure and evaluate the probabilities of electron-positron annihilation
in positronium hydrides. In this Section we restrict ourselves to the
analysis of the two systems: ${}^{\infty}$HPs and ${}^{1}$HPs. The total
energies obtained for the ground state of these two systems in our
calculations are -0.789 196 764 445 $a.u.$ and -0.788 870 709 151 $a.u.$,
respectively. The proton mass used in these calculations is $M_p$ =
1836.15267261 $m_e$. For masses of deuterium and tritium nuclei we have used
the values $M_d$ = 3670.4829652 $m_e$ and $M_t$ = 5496.92158 $m_e$.

The absolute value of the hyperfine structure splitting in the ${}^{1}$HPs
system is $\approx 3.61$ $MHz$. This value represents the energy splitting
between the triplet and singlet states of hyperfine structure. As expected
the absolute value of hyperfine structure splitting in the ${}^{1}$HPs
system is significantly smaller ($\approx$ 6 times smaller) than such a
splitting in the MuPs system (see Section II). In the DPs and TPs hydrides
the hyperfine structure splittings are also relatively small.

Annihilation rates for the ${}^{\infty}$HPs system are: $\Gamma_{2 \gamma}
\approx 2.4568468(30) \cdot 10^9$ $sec^{-1}$, $\Gamma_{3 \gamma} \approx
6.656047(10) \cdot 10^6$ $sec^{-1}$, $\Gamma_{4 \gamma} \approx 3.6321(1)
\cdot 10^3$ $sec^{-1}$ and $\Gamma_{5 \gamma} \approx 6.3565(1)$ $sec^{-1}$,
respectively. The first one-photon annihilation rate in the ${}^{\infty}$HPs
system is $\Gamma^{(1)}_{1 \gamma} \approx$ 3.93809(10)$\cdot 10^{-1}$ $%
sec^{-1}$, while the zero-photon annihilation rate is $\Gamma_{0 \gamma}
\approx \xi_0 \cdot 9.4502(1) \cdot 10^{-8}$ $sec^{-1}$, where $\xi_0$ is an
unknown numerical factor ($\xi_0 \approx 1$). For the ${}^{1}$HPs system
these annihilation rates are: $\Gamma_{2 \gamma} \approx 2.4562527(30) \cdot
10^9$ $sec^{-1}$, $\Gamma_{3 \gamma} \approx 6.654438(10) \cdot 10^6$ $%
sec^{-1}$, $\Gamma_{4 \gamma} \approx 3.6312(1) \cdot 10^3$ $sec^{-1}$ and $%
\Gamma_{5 \gamma} \approx 6.3550(1)$ $sec^{-1}$, respectively. The first
one-photon annihilation rate is $\Gamma^{(1)}_{1 \gamma} \approx$ 3.93025(10)%
$\cdot 10^{-1}$ $sec^{-1}$ and zero-photon annihilation rate is $\Gamma_{0
\gamma} \approx \xi_1 \cdot 9.4289(1) \cdot 10^{-8}$ $sec^{-1}$, where $\xi_1
$ is an unknown numerical factor. No attempt was made to evaluate the second
one-photon annihilation rate $\Gamma^{(2)}_{1 \gamma}$ accurately in this
study. Annihilation rates for the DPs and TPs hydrides can be found in Table
I.

It should be mentioned that our current expressions for annihilation rates $%
\Gamma_{2 \gamma} \Gamma_{3 \gamma}, \Gamma_{4 \gamma}$ and $\Gamma_{5
\gamma}$ which are used above for MuPs and other positronium hydrides have
been derived from a rigorous consideration based on Quantum Electrodynamics
whereas the annihilation rates determined in Ref.\cite{FrSm} are based on
very approximate relations.

\section{Muonium-positronium conversion.}

In the four-body MuPs system there is a possibility to observe a very
interesting process of muonium (or $\mu^{+} e^{-}$) conversion into the
charge conjugate system $\mu^{-} e^{+}$ \cite{Litt} - \cite{FroX}. The
muonium-antimuonium conversion has attracted significant theoretical and
experimental attention for many years (see, e.g., \cite{Litt} - \cite{FroX}, 
\cite{Pont}, \cite{Fein} and references therein). In atomic physics such a
process corresponds to a spontaneous conversion of the incident atom into
its anti-atom. In the four-body MuPs system this process is even more
interesting, since during such a conversion the original $\mu^{+} e^{-}_2
e^{+}$ system is transformed into the charge conjugate four-body system $%
\mu^{-} e^{+}_2 e^{-}$ (or $\overline{\mathrm{Mu}}$Ps) in which the heaviest
particle has the negative charge. The newly arising $\mu^{-} e^{+}_2 e^{-}$
system contains two positrons $e^{+}$ and one electron $e^{-}$. Very likely
that the newly arising system $\overline{\mathrm{Mu}}$Ps can be in the same
(atomic) bound state as the original system MuPs.

Formally, the muonium-positronium conversion is not prohibited by any
conservation law. However, it is very hard to observe such a conversion
under actual experimental conditions. Mainly, this is related to the very
short life-time of the incident MuPs system. The positively charged muon $%
\mu^+$ is an unstable particle which decays as follows: 
\begin{eqnarray}
\mu^+ \rightarrow e^+ + \nu_e + \overline{\nu}_{\mu}  \label{dec}
\end{eqnarray}
where $\nu_e$ and $\overline{\nu}_{\mu}$ are the electron neutrino and muon
antineutrino, respectively. The muon mean life-time $\tau_{\mu}$ is $\approx
(2.19703 \pm 4 \cdot 10^{-5}) \cdot 10^{-6}$ $sec$. That part of the Fermi
theory Lagrangian $\mathcal{L}_F$ which corresponds to the muon decay Eq.(%
\ref{dec}) is 
\begin{eqnarray}
\mathcal{L}_W = - \frac{1}{\sqrt{2}} G_F \bigl[ \overline{\psi}_{\nu_{\mu}}
\gamma_{\lambda} (1 + \gamma_5) \psi_{\mu} \bigr] \bigl[ \overline{\psi}_e
\gamma_{\lambda} (1 + \gamma_5) \psi_{\nu_{e}} \bigr] \; \; \; .
\end{eqnarray}
where $G_F$ is the Fermi coupling constant, while $\psi_{\mu}, \psi_e,
\psi_{\nu_{e}}$ and $\psi_{\nu_{\mu}}$ are the wave functions of the muon,
electron, electron neutrino and muon antineutrino, respectively. Also in
this equation $\gamma_{\lambda}$ and $\gamma_5 = \imath \gamma_1 \gamma_2
\gamma_3 \gamma_0$ are the corresponding Dirac ($4 \times 4$) matrices.

In general, the Fermi theory Lagrangian $\mathcal{L}_F$ must also include
the bare quantum-electrodynamic Lagrangian $\mathcal{L}_{QED}$ and bare
quantum chromodynamic Lagrangian $\mathcal{L}_{QCD}$ which is responsible
for strong interactions, i.e. $\mathcal{L}_F = \mathcal{L}_W + \mathcal{L}%
_{QED} + \mathcal{L}_{QCD}$. The $\mathcal{L}_{QCD}$ Lagrangian is not of
interest for our present purposes. The quantum-electrodynamic Lagrangian $%
\mathcal{L}_{QED}$ is of the form 
\begin{eqnarray}
\mathcal{L}_{QED} = - \sum_f \overline{\psi}_f (\imath \gamma_{\lambda}
p_{\lambda} + m_f) \psi_f - \frac14 (\partial_{\kappa} A_{\lambda} -
\partial_{\lambda} A_{\kappa})^2 + \imath e \sum_f Q_f (\overline{\psi}_f
\gamma_{\lambda} \psi_f) A_{\lambda} \; \; \; ,  \label{two}
\end{eqnarray}
where the explicitly shown sums are over all fermion species $f$ (in the
present case $f = \mu^+, e^+$), with rest mass $m_f$ and electric charge $Q_f
$ (in the units of $e$). The notation $A_{\lambda}$ stands for the four
components of the electromagnetic field $A_{\lambda} = (A_0, -\mathbf{A})$.
The Greek letters $\kappa$ and $\lambda$ designate four-dimensional indices,
taking on the values 0, 1, 2, 3. In these equations and Eq.(\ref{conn})
below the sum is assumed over any repeated Greek index and the summation
sign will not be used in such cases.

The analytical expression for the decay rate $\Gamma_{\mu}$ of positive muon 
$\mu^+$ follows from the Fermi $V - A$ theory \cite{Rit} 
\begin{eqnarray}
\Gamma_{\mu} = \frac{1}{\tau_{\mu}} = \frac{G^2_F \cdot m^5_{\mu}}{192 \pi^3}
(1 + \Delta q) \; \; \; ,  \label{e45}
\end{eqnarray}
where $G_F$ is the Fermi coupling constant, $m_{\mu}$ is the muon rest mass
and $\Delta q$ is the corresponding relativistic correction \cite{Rit}. The
current value of the Fermi constant $G_F$ is ($1.16637 \pm 0.00001) \cdot
10^{-5}$ $GeV^{-2}$ \cite{Rit} (see also \cite{Bard}, \cite{Fets} and
references therein).

In general, the branching ratio of the muonium conversion is determined by
the ratio $R_g$ of the conversion $G_C$ and Fermi coupling constants $G_F$,
i.e. $R_g = \frac{G_C}{G_F}$. The conversion constant $G_C$ appears in the
effective Lagrangian $\mathcal{L}_C$ for the $\mu^+e^- \rightarrow \mu^-e^+$
conversion 
\begin{eqnarray}
\mathcal{L}_C = \frac{1}{\sqrt{2}} G_C \bigl[ \overline{\psi}_{\mu}
\gamma_{\lambda} (1 - \gamma_5) \psi_e \bigr] \bigl[ \overline{\psi}_{\mu}
\gamma_{\lambda} (1 - \gamma_5) \psi_e \bigr] + h.c.  \label{conn}
\end{eqnarray}
where `$h.c.$' means the hermitian conjugate expression. The theoretically
predicted ratio $R_g = \frac{G_C}{G_F}$ is relatively small (see, e.g., \cite%
{Pont}). The total probability of the $\mu^+e^- \rightarrow \mu^-e^+$
conversion can be approximately represented in the form \cite{Pont}, \cite%
{Fein} 
\begin{eqnarray}
P_c \approx 2.6 \cdot 10^{-5} \cdot \Bigl( \frac{G_C}{G_F} \Bigr)^2 \approx
2.34 \cdot 10^{-10} \; \; \; ,  \label{eq22}
\end{eqnarray}
where we used the most recent experimental value of the $R_g$ ratio $R_g
\approx 0.0030$ \cite{Will}. In fact, in \cite{Will} it was found that $R_g
< 0.0030$. The results of other experiments in which $R_g$ has been measured
at different energies can be found in \cite{Gord} ($R_g \le 0.14$) and in 
\cite{Abel} ($R_g \le 0.008$). The evaluation which follows from Eq.(\ref%
{eq22}) indicates that we can observe muonium conversion only in two MuPs
systems of each 100 millions created in experiments.

\section{Conclusion.}

The hyperfine structure splitting and annihilation of the electron-positron
pairs in the ground bound state of muonium-positronium MuPs has been
studied. Its is shown that the hyperfine splitting between singlet $J = 0$
and triplet $J = 1$ spin states in MuPs is $\approx$ 22.958(10) $MHz$. We
also consider the annihilation of electron-positron pairs in the MuPs
system. The largest two-photon annihilation rate is $\Gamma_{2 \gamma}
\approx 2.4522354(30) \cdot 10^9$ $sec^{-1}$. The numerical values of the
three-, four- and five-photon annihilations are $\Gamma_{3 \gamma} \approx
6.643554(10) \cdot 10^6$ $sec^{-1}$, $\Gamma_{4 \gamma} \approx 3.6253(1)
\cdot 10^3$ $sec^{-1}$ and $\Gamma_{5 \gamma} \approx 6.3446(1)$ $sec^{-1}$,
respectively. These values are accurate and based on the results of rigorous
QED analysis, rather than on approximate relations used in our earlier work 
\cite{FrSm}. The rates of zero- and one-photon annihilations have been also
determined for the MuPs system: $\Gamma_{0 \gamma}$(MuPs) $\approx \xi \cdot
9.1032(1) \cdot 10^{-8}$ $sec^{-1}$ and $\Gamma^{(1)}_{1 \gamma}$(MuPs) $%
\approx 3.8706(1) \cdot 10^{-1}$ $sec^{-1}$. The second one-photon
annihilation rate $\Gamma_{1 \gamma}^{(2)}$(MuPs) has not been evaluated in
this study. The expression for zero-photon annihilation rate $\Gamma_{0
\gamma}$(MuPs) also contains an unknown numerical factor $\xi$ which must be
derived from Quantum Electrodynamics.

Analogous annihilation rates have been evaluated for other positronium
hydrides ${}^{\infty}$HPs, TPs, DPs and ${}^{1}$HPs. Note that in our
current computations we have used the variational expansions based on
six-dimensional gaussoids \cite{KT} in which all non-linear parameters have
been varied. The most recent version of this method includes a number of
substantial improvements made in the optimization of the non-linear
parameters and in overall accuracy and numerical stability of our procedure 
\cite{BaFr}, \cite{Fro09}. Finally, we improved the results of previous
studies performed for positronium hydrides. We also discuss the possibility
to observe the muonium-antimuonium conversion in MuPs. It is shown that such
a conversion transforms the incident four-body system $\mu^{+} e^{-}_2 e^{+}$
into its charge conjugate system $\mu^{-} e^{-} e^{+}_2$. It is expected
that the new system $\mu^{-} e^{-} e^{+}_2$ will remain in the same bound
state (the ground $1^1S_e$-state).


\begin{thebibliography}{99}
\bibitem{Ore} A. Ore, Phys. Rev. \textbf{83}, 665 (1951).

\bibitem{Dra1} R.J. Drachman, \textit{Positron Astrophysics}, in: Positron
Annihilation, edited by P.G. Coleman, S.C. Sharma and L.M. Diana
(North-Holland, Amsterdam, 1982), p. 37 - 42 and references therein.

\bibitem{Dra2} R.J. Drachman, Can. J. Phys. \textbf{60}, 494 (1982).

\bibitem{HPs} D.M. Schrader, F.M. Jacobsen, N.-P. Fradsen and U. Mikkelsen,
Phys. Rev. Lett. \textbf{69}, 57 (1992).

\bibitem{Hou} S.K. Houston and R.J. Drachman, Phys. Rev. A \textbf{7}, 819
(1973).

\bibitem{Nav} P.B. Navin, D.M. Schrader and C.F. Lebeda, Phys. Rev. A 
\textbf{9}, 2248 (1974).

\bibitem{Page} B.A. Page and P.A. Frazer, J. Phys. B \textbf{7}, L389 (1974).

\bibitem{Ho1} Y.K. Ho, Phys. Rev. A \textbf{34}, 609 (1986).

\bibitem{Ho2} Y.K. Ho, Phys. Rev. A \textbf{48}, 4780 (1993).

\bibitem{Sch} D.M. Schrader, \textit{Chemical Stability and Approximate
Quantum Mechanics}, in: Positron Annihilation, edited by P.G. Coleman, S.C.
Sharma and L.M. Diana (North-Holland, Amsterdam, 1982), p. 37 - 42 and
references therein.

\bibitem{Fro1} A.M. Frolov, Phys. Rev. A \textbf{69}, 062507 (2004).

\bibitem{Bubin} S. Bubin and L. Adamowicz, Phys. Rev. A \textbf{74}, 052502
(2006).

\bibitem{FrSm} A.M. Frolov and V.H. Smith, Jr., Phys. Rev. A \textbf{55},
2662 (1997).

\bibitem{NIST} The NIST Reference on Constants, Units and Uncertainty,
\linebreak see: http://physics.nist.gov/cuu/Constants/index.html

\bibitem{CRC} \textit{CRC Handbook of Chemistry and Physics}, 85th Edition,
Ed. D.R. Lide, (CRC Press, Inc., Boca Raton, Florida, 2004).

\bibitem{AB} A.I. Akhiezer and V.B. Beresteskii, \textit{Quantum
Electrodynamics}, (4th Ed., Nauka (Science), Moscow (1981)), Chps. 4 and 5
(in Russian).

\bibitem{Harr} I. Harris and L.M. Brown, Phys. Rev. \textbf{105}, 1656
(1957).

\bibitem{Heitl} W. Heitler, \textit{The Quantum Theory of Radiation}, (3rd.
Ed., Oxford at the Clarendon Press, Oxford (UK), 1954), Chp. V.

\bibitem{PRA83} G.P. Lepage, P.B. Mackenzie, K.H. Streng and P.M. Zerwas,
Phys. Rev. \textbf{28}, 3090 (1983).

\bibitem{Kru} A.M. Frolov, S.I. Kryuchkov and V.H. Smith, Jr., Phys. Rev. A 
\textbf{51}, 3636 (1995).

\bibitem{Kry} S.I. Kryuchkov, J. Phys. B \textbf{27}, L61 (1994).

\bibitem{Litt} L.S. Littenberg and R. Shrock, Phys. Lett. B \textbf{491},
285 (2000).

\bibitem{WHou} W.-S. Hou and G.G. Wong, Phys. Lett. B \textbf{357}, 145
(1995).

\bibitem{Hub} T.M. Hubert et al, Phys. Rev. D \textbf{41}, 2709 (1990).

\bibitem{Aoki} M. Aoki, Muonium to Anti-Muonium Convesrion and $\mu^{-} -
\mu^{+}$ conversion, unpublished (2001).

\bibitem{FroX} A.M. Frolov, J. Phys. B \textbf{37}, 2191 (2004).

\bibitem{Pont} B. Pontecorvo, Zh. Eksp. Teor. Fiz. \textbf{33}, 549 (1957)
[Sov. Phys. JETP \textbf{6}, 429 (1958)].

\bibitem{Fein} G. Feinberg and L.M. Lederman, Ann. Rev. Nucl. Sci. \textbf{13%
}, 431 (1961).

\bibitem{Rit} T. van Ritbergen and R.G. Stuart, Phys. Rev. Lett. \textbf{82}%
, 488 (1999).

\bibitem{Bard} G. Bardin et al, Phys. Lett. B \textbf{137}, 135 (1984).

\bibitem{Fets} W. Fetscher, H.-J. Gerber and K.F. Johnson, Phys. Lett. B 
\textbf{173}, 102 (1986).

\bibitem{Will} L. Willmann et al, Phys. Rev. Lett. \textbf{82}, 49 (1999).

\bibitem{Gord} V.A. Gordeev et al, Yad. Fiz. \textbf{60}, 1291 (1997) [Phys.
At. Nuclei \textbf{60}, 1164 (1997)].

\bibitem{Abel} R. Abela et al, Phys. Rev. Lett. \textbf{77}, 1950 (1996).

\bibitem{KT} N.N. Kolesnikov and V.I. Tarasov, Yad. Fiz. \textbf{35}, 609
(1982), [Sov. J. Nucl. Phys. \textbf{35}, 354 (1982)].

\bibitem{BaFr} D.H. Bailey and A.M. Frolov, Phys. Rev. A \textbf{72}, 014501
(2005).

\bibitem{Fro09} A.M. Frolov and D.M. Wardlaw, Jurn. Exp. Teor. Fiz. \textbf{%
135}, 667 (2009) [JETP \textbf{108}, 583 (2009)].
\end{thebibliography}
\end{document}